\documentclass[preprint,aps,showpacs]{revtex4}
\usepackage{amsfonts}
\usepackage{amsmath}
\usepackage{amssymb}
\usepackage{graphicx}

\begin{document}

\title{Dynamic $0-\pi$ transition induced by pumping mechanism}
\author{Jun Wen, Baigeng Wang, Qingyun Zhang, D. Y. Xing}
\affiliation{National Laboratory of Solid State Microstructures and Department of
Physics, Nanjing University, Nanjing 210093, China}
\date{\today}

\begin{abstract}
Using Nambu$\otimes $spin space Keldysh Green's function approach, we
present a nonequilibrium charge and spin pumping theory of a quantum dot in
the mico-cavity coupled to two superconducting leads. It is found that the
charge currents include two parts: The dissipationless supercurrent standing
for the transfer of coherent Cooper pairs and the pumped quasi-particle
current. The supercurrent exhibits a dynamic $0-\pi $ transition induced by
the frequency and strength of the $\sigma _{-}$ polarized laser field. This
dynamic transition is not affected by the strong Coulomb interaction.
Especially, the spin current appears and is an even function of the phase
difference between two superconductors when the frequency of the polarized
laser field is larger than two times superconducting energy gap. Our theory
serves as an extension to non-superconducting spintronics.
\end{abstract}

\maketitle

The interplay between superconducting order and ferromagnetic order has
attracted considerable attention recently in the proximity effects near the
interface of ferromagnet/superconductor heterostructures, since it subsumes
many fascinating physical phenomena at the interface and leads to potential
applications that may complement non-superconducting spintronic devices (see
Reviews\cite{golubov,bergeret,buzdin} and Refs. therein). A interesting
physical phenomenon due to the interplay of magnetic and superconducting
orders is the so-called $0-\pi $ transition in
superconductor-ferromagnet-superconductor (S/F/S) junctions\cite{bulaevskii,
buzdin1,oboznov,born,robinson,andersen} and the magnetic quantum dots\cite%
{kowenhowen,arovas}. The existence of the $\pi $ junction in layered S/F/S
systems was first predicted by Bulaevskii \textit{et al.}\cite{bulaevskii}
and Buzdin \textit{et al.}\cite{buzdin1} to occur for certain thicknesses
and exchange field energies of the F layer and was later confirmed in the
experiment\cite{oboznov}. These investigations are not only of academic
interest, but also of importance for the solid-state implementation of a
quantum bit based on a superconducting loop with $0$ and $\pi $ Josephson
junctions\cite{many1}.

The present work is also related to the parametric quantum pumping\cite%
{brouwer}, which refers to the transfer of electrons{\normalsize \ }%
coherently between two reservoirs at zero bias. In particular, when the
charge transfer is quantized per cycle, the pumping could be of importance
in establishing a standard of electric current\cite{many2}. Therefore,
charge pumping has attracted much attention since the first proposal by
Thouless\cite{thouless}. The first experimental demonstration of the pumping
of charges was reported by Switkes \textit{et al.}\cite{switkes}. They
applied sinusoidal gate voltages to an open quantum dot to pump the charge
current. More recently, this pumping mechanism has been extended to
different transport contexts, including the generation of spin current\cite%
{marcus,bgwang,jwang,benjamin,aono} and entangled pairs\cite%
{buttikers,beenakker}.

In this Letter, we present a nonequilibrium charge and spin transport theory
of a quantum dot under the optical micro-cavity. We find that the laser
field acting as the pumping forces can induce the dynamic $0-\pi $
transition which is not affected by the strong Coulomb interaction. Our
theory goes beyond the adiabatic limit, and is valid for the arbitrary
pumping frequency and temperature. More interestingly, we find the spin
current as an even function of the phase difference between two
superconductors is pumped when the laser frequency is larger than $2\Delta $.

The system we consider is a quantum dot embedded in a high-Q micro-cavity.
Two superconducting reservoirs are coupled to the dot via tunneling. The
Hamiltonian of the present system reads%
\begin{equation}
H=H_{0}+H(t),
\end{equation}%
in which the time-independent $H_{0}$ consists of three parts: $H_{S}$
describes the left and right ordinary BCS superconductors with the energy
gap function $\Delta _{L,R}$ and band width $W$%
\begin{equation}
H_{S}=\sum_{k\sigma \alpha =L,R}\epsilon _{k}c_{k\sigma \alpha }^{\dagger
}c_{k\sigma \alpha }+\sum_{k\alpha =L,R}[\Delta _{\alpha }c_{k\uparrow
\alpha }^{\dagger }c_{-k\downarrow \alpha }^{\dagger }+h.c.].
\end{equation}%
We assume $\Delta _{\alpha }=\Delta \exp (i\Phi _{\alpha })$, \textit{i.e.},
two superconductors have the same energy gap but the different phase. $H_{D}$
stands for the Hamiltonian of single energy level quantum dot with the
on-site Coulomb interaction%
\begin{equation}
H_{D}=\sum_{\sigma }\epsilon _{\sigma }d_{\sigma }^{\dagger }d_{\sigma
}+Un_{\uparrow }n_{\downarrow }.
\end{equation}%
Note that the Zeeman splitting due to the external magnetic field $B$
between two spin states is $h\equiv \epsilon _{\uparrow }-\epsilon
_{\downarrow }=g\mu _{B}B$, and $U$ is on-site Coulomb interaction. $H_{T}$
is the tunneling Hamiltonian between the superconductors and the quantum dot%
\begin{equation}
H_{T}=\sum_{k\sigma \alpha }[T_{k\sigma \alpha }c_{k\sigma \alpha }^{\dagger
}d_{\sigma }+T_{k\sigma \alpha }^{\ast }d_{\sigma }^{\dagger }c_{k\sigma
\alpha }].
\end{equation}%
The most important term $H(t)$ denotes transitions between the different
spin states of the dot, which can be induced by a two-photo Raman process%
\cite{search}. If we treat the strong laser field classically, $H(t)$ has
the following form%
\begin{equation}
H(t)=r[d_{\uparrow }^{\dagger }d_{\downarrow }\exp (-i\omega t)+h.c.].
\end{equation}%
Here $r$ and $\omega $ are the classical Rabi frequency and the $\sigma _{-}$
polarized laser frequency, respectively. This $\sigma _{-}$ polarized laser
acts as the two pumping forces with the phase difference $\pi /2.$ It must
be pointed out that neglecting the on-site Coulomb interaction and replacing
the superconductor by normal metal in the above model have been used to
generate spin current\cite{bgwang,jwang,search}. Due to the existence of
spin-flip term, we have to work in the generalized Nambu$\otimes $spin
space. The standard Keldysh Green's function gives the charge and spin
current $(e=\hbar =1)$%
\begin{eqnarray}
I_{c,s}(t) &=&\int \frac{dE_{1}}{2\pi }\frac{dE_{2}}{2\pi }Tr\{\sigma
_{c,s}[G^{r}(E_{1},E_{2})\Sigma _{L}^{<}(E_{2})+G^{<}(E_{1},E_{2})\Sigma
_{L}^{a}(E_{2})+h.c.]\}  \notag \\
&&\ast \exp [-i(E_{1}-E_{2})t]
\end{eqnarray}%
where $\sigma _{c}\equiv \frac{1}{2}\left(
\begin{tabular}{cc}
$\sigma _{z}$ & $0$ \\
$0$ & $\sigma _{z}$%
\end{tabular}%
\right) ,$ $\sigma _{s}\equiv \frac{1}{4}\left(
\begin{tabular}{cc}
$\hat{1}$ & $0$ \\
$0$ & $-\hat{1}$%
\end{tabular}%
\right) $. $G^{r}(E_{1},E_{2})$ and $G^{<}(E_{1},E_{2})$ are the Fourier
representations of the retarded and lesser Green's functions in the
generalized Nambu$\otimes $spin space with the creation operator $\Psi
^{\dagger }=(d_{\uparrow }^{\dagger },d_{\downarrow },d_{\downarrow
}^{\dagger },d_{\uparrow })$. They have the usual form $G^{r}(t_{1},t_{2})%
\equiv -i\theta (t_{1}-t_{2})\langle \{\Psi (t_{1}),\Psi ^{\dagger
}(t_{2})\}\rangle $ and $G^{<}(t_{1},t_{2})\equiv i\langle \Psi ^{\dagger
}(t_{2})\Psi (t_{1})\rangle ,$respectively. $\Sigma _{\alpha }^{r,<}(E)$ are
the retarded and lesser self-energies due to the superconducting lead $%
\alpha $. \ The lesser Green's function is related to the retarded and
advanced Green's functions via Keldysh equation $G^{<}=$ $G^{r}\Sigma
^{<}G^{a}$, therefore, the whole problem is reduced to the calculation of
the retarded Green's function $G^{r}$. This can be done by two steps: we
first calculate the time-independent retarded Green's function $G_{0}^{r}(E)$
of the quantum dot, then the full Green's function can be obtained exactly
from the Dyson equation $G^{r}(t_{1},t_{2})=G_{0}^{r}(t_{1}-t_{2})+\int
dtG_{0}^{r}(t_{1}-t)H(t)G^{r}(t,t_{2}),$which has the following Fourier form%
\begin{equation}
\left(
\begin{tabular}{ll}
$G_{2i}^{r}(E_{1},E_{2})$ & $G_{1i}^{r}(E_{1},E_{2})$%
\end{tabular}%
\right) =\left(
\begin{tabular}{ll}
$C_{2i}$ & $C_{1i}$%
\end{tabular}%
\right) \hat{A}/Det\hat{A},
\end{equation}

\begin{equation}
\left(
\begin{tabular}{ll}
$G_{4i}^{r}(E_{1},E_{2})$ & $G_{3i}^{r}(E_{1},E_{2})$%
\end{tabular}%
\right) =\left(
\begin{tabular}{ll}
$C_{4i}$ & $C_{3i}$%
\end{tabular}%
\right) \hat{B}/Det\hat{B},
\end{equation}%
\ where the matrices $\hat{A}$, $\hat{B}$ and the coefficients $C_{ij}$ are
defined as%
\begin{equation}
\hat{A}=\hat{1}+r^{2}\left(
\begin{tabular}{cc}
$-G_{011}^{r}(E_{1})$ & $G_{012}^{r}(E_{1})$ \\
$G_{021}^{r}(E_{1})$ & $-G_{022}^{r}(E_{1})$%
\end{tabular}%
\right) \left(
\begin{tabular}{cc}
$G_{033}^{r}(E_{1}-\omega )$ & $G_{034}^{r}(E_{1}-\omega )$ \\
$G_{043}^{r}(E_{1}-\omega )$ & $G_{044}^{r}(E_{1}-\omega )$%
\end{tabular}%
\right) ,
\end{equation}%
\begin{equation}
\hat{B}=\hat{1}+r^{2}\left(
\begin{tabular}{cc}
$-G_{033}^{r}(E_{1})$ & $G_{034}^{r}(E_{1})$ \\
$G_{043}^{r}(E_{1})$ & $-G_{044}^{r}(E_{1})$%
\end{tabular}%
\right) \left(
\begin{tabular}{cc}
$G_{011}^{r}(E_{1}+\omega )$ & $G_{012}^{r}(E_{1}+\omega )$ \\
$G_{021}^{r}(E_{1}+\omega )$ & $G_{022}^{r}(E_{1}+\omega )$%
\end{tabular}%
\right) ,
\end{equation}%
\begin{equation}
C_{ij}=\left\{
\begin{tabular}{l}
$2\pi G_{0ij}^{r}(E_{1})\delta (E_{1}-E_{2})$ \\
$if$ $i,j=1,2$ $or$ $i,j=3,4$ \\
$2\pi r[G_{0i1}^{r}(E_{1})G_{03j}^{r}(E_{1}-\omega
)-G_{0i2}^{r}(E_{1})G_{04j}^{r}(E_{1}-\omega )]\delta (E_{1}-E_{2}-\omega )$
\\
$if$ $i=1,2$ $and$ $j=3,4$ \\
$2\pi r[G_{0i3}^{r}(E_{1})G_{01j}^{r}(E_{1}+\omega
)-G_{0i4}^{r}(E_{1})G_{02j}^{r}(E_{1}+\omega )]\delta (E_{1}-E_{2}+\omega )$
\\
$if$ $i=3,4$ $and$ $j=1,2$%
\end{tabular}%
\right. .
\end{equation}%
Substituting the retarded Green's function into Eq.(6), we can obtain the
charge and spin current formulae directly, which have the following form

\begin{equation}
I_{c}=\int \frac{dE}{2\pi }\mathrm{Tr}\{\sigma _{c}[\hat{F}(E)f+\hat{F}%
^{+}(E)(f^{+}-f)+\hat{F}^{-}(E)(f^{-}-f)+h.c.]\},
\end{equation}%
\begin{equation}
I_{s}=\int \frac{dE}{2\pi }\mathrm{Tr}\{\sigma _{s}[\hat{F}^{+}(E)(f^{+}-f)+%
\hat{F}^{-}(E)(f^{-}-f)+h.c.]\}.
\end{equation}%
Here $\hat{F}(E)=G^{r}(E,E)[\Sigma _{L}^{a}(E)-\Sigma
_{L}^{r}(E)]+\int \frac{dX}{2\pi }\{G^{r}(E,X)[\Sigma ^{a}(X)-\Sigma
^{r}(X)]G^{a}(X,E)\Sigma _{L}^{a}(E)\}$, $\hat{F}^{+}(E)=\int
\frac{dX}{2\pi }\{G^{r}(E,X)\tau _{1}[\Sigma ^{a}(X)-\Sigma
^{r}(X)]\tau _{1}G^{a}(X,E)\tau _{2}\Sigma _{L}^{a}(E)\tau _{2}\}$,
$\hat{F}^{-}(E)=\int \frac{dX}{2\pi }\{G^{r}(E,X)\tau _{2}[\Sigma
^{a}(X)-\Sigma ^{r}(X)]\tau _{2}G^{a}(X,E)\tau _{1}\Sigma
_{L}^{a}(E)\tau _{1}\}$\cite{footnote1}. The matrices $\tau _{1}$
and $\tau _{2}$ are defined as $\tau _{1}=$ $\left(
\begin{tabular}{ll}
$\hat{1}$ & $0$ \\
$0$ & $0$%
\end{tabular}%
\right) $, $\tau _{2}=$ $\left(
\begin{tabular}{ll}
$0$ & $0$ \\
$0$ & $\hat{1}$%
\end{tabular}%
\right) $respectively.$\ f$ is the well-known Fermi distribution function $%
f\equiv 1/(\exp (\beta E)+1)$ and $f^{\pm }\equiv f(E\pm \omega )$. Eq.(12)
shows that the charge current comes from two parts: the dissipationless
supercurrent $I_{sc}=\int \frac{dE}{2\pi }\mathrm{Tr}\{\sigma _{c}[\hat{F}%
(E)f+h.c.]\}$ involving in the transfer of coherent Cooper pairs and
dissipative quasi-particle current $I_{qc}=\int \frac{dE}{2\pi }\mathrm{Tr}%
\{\sigma _{c}[\hat{F}^{+}(E)(f^{+}-f)+\hat{F}^{-}(E)(f^{-}-f)+h.c.]\}$.
Since the Cooper pair is singlet, only can the spin current be carried by
the dissipative quasi-particle process. We must stress that the above charge
and spin current formulae go beyond the adiabatic approximation, and are
valid for the arbitrary temperature and frequency and strength of the laser
field since we can calculate the retarded Green's function exactly.

We first consider the non-interacting case on the quantum dot. The
time-independent retarded Green's function for the quantum dot is given by%
\begin{equation}
G_{0}^{r}(E)=\{E\hat{1}-H_{D}-\sum_{\alpha }\Sigma _{\alpha }^{r}(E)\}^{-1},
\end{equation}%
where $H_{D}\equiv \left(
\begin{tabular}{cccc}
$\epsilon _{\uparrow }$ & $0$ & $0$ & $0$ \\
$0$ & $-\epsilon _{\downarrow }$ & $0$ & $0$ \\
$0$ & $0$ & $\epsilon _{\downarrow }$ & $0$ \\
$0$ & $0$ & $0$ & $-\epsilon _{\uparrow }$%
\end{tabular}%
\right) $, $\Sigma _{\alpha }^{r}(E)\equiv -\frac{i\Gamma _{\alpha }\xi _{E}%
}{2\sqrt{E^{2}-\Delta ^{2}}}\left(
\begin{tabular}{cccc}
$E$ & $-\Delta _{\alpha }$ & $0$ & $0$ \\
$-\Delta _{\alpha }^{\ast }$ & $E$ & $0$ & $0$ \\
$0$ & $0$ & $E$ & $\Delta _{\alpha }$ \\
$0$ & $0$ & $\Delta _{\alpha }^{\ast }$ & $E$%
\end{tabular}%
\right) $ with $\xi _{E}=1$ for $E>-\Delta $ and $\xi _{E}=-1$ otherwise. $%
\Gamma _{\alpha }$ is the linewidth function which is assumed to be a
constant under wide-band limit. Once having the above time-independent
retarded Green's function, we can obtain the full Green's function and then
charge and spin current by using Eqs.(7-13). Fig.1(a) shows the Josephson
current $I_{sc}$ versus the two superconducting phase difference $\Phi
\equiv \Phi _{R}-\Phi _{L}$ of two superconductors for the various polarized
laser frequencies $\omega .$ Our system is in the $\pi $-state when the
laser frequency is lower. As the frequency $\omega $ increases, the $\pi $
state will turn into a $0$-state first, \textit{i.e.}, a dynamic $\pi -0$
transition happens. For a higher frequency the $0$-state will experience a
dynamic $0-\pi $ transition and become a $\pi $-state again. This result can
be seen clearly in Fig.2, which exhibits the Josephson current $I_{sc}$ as a
function of frequency $\omega .$ In Fig.1(b) we plot the Josephson current $%
I_{sc}$ versus the phase difference $\Phi $ of two superconductors for the
different polarized laser strength $r.$ The similar $0-\pi $ transition can
be modulated by the strength $r$ of laser field. Fig.3 describes the pumped
quasi-particle charge current $I_{qc}$ versus the laser frequency. We find
that this pumped quasi-particle charge current is very small for the lower
frequency and increases with frequency non-monotonously. The pumped
quasi-particle charge current can be understood from the co-tunneling
process. For example, a spin down electron $1$ in the left superconductor
tunnels into the quantum dot, then a Cooper pair splits into two electrons $%
2 $ and $2^{\prime }$ with the opposite spin. The spin down electron $%
2^{\prime }$ fills in the original position of the electron $1$, while spin
up electron $2$ will tunnel into the quantum dot. Finally, the spin down
electron $1$ and\ spin up electron $2$ in the quantum dot change their spin
by absorption and emitting a photon, respectively, and form a pair to enter
the right superconductor. This co-tunneling process is depicted in the inset
of Fig.3. Our numerical calculation shows that the pumped quasi-particle
charge current always satisfies $I_{qc11}=$ $I_{qc22}=$ $I_{qc33}=$ $%
I_{qc44} $ and gives no spin current if only the laser frequency is lower
than $2\Delta $. However, the pumped spin current appears once the laser
frequency is higher than $2\Delta .$ The result can be seen from Fig.4.
Obviously, this spin current arises from the additional spin-flip process
which is plotted in the inset of Fig.4: A spin-down electron from both
superconducting leads can tunnel into the quantum dot, and due to the $%
\sigma _{-}$ polarized laser field it absorbs a photon and becomes a spin-up
electron. If the laser frequency is higher than $2\Delta $, this spin-up
electron can tunnel out of the scattering region and goes into the
superconducting leads as a quasi-particle. In the Fig.1(c) we plot the spin
current $I_{s}$ versus the two superconducting phase difference $\Phi $. In
contrast to the Josephson supercurrent, the pumped spin current $I_{s}$ is
an even function of $\Phi $.

Finally, we briefly consider there exists a strong Coulomb interacting case $%
(U\rightarrow \infty )$ in the quantum dot. We focus on the strong coupling
limit $\Delta <<T_{K}$, where the Kondo effect becomes important and leads
to a positive pairing correlation on the quantum dot. This situation is well
described in the slave-boson language. The physical electron operator $%
d_{\sigma }$ can be replaced by $b^{+}f_{\sigma }$ , where $b$ and $%
f_{\sigma }$ being the standard boson and fermion annihilation operators
standing for the empty $(n_{\uparrow }=0,n_{\downarrow }=0)$ and singly
occupied $(n_{\uparrow }=1,n_{\downarrow }=0)$ or $(n_{\uparrow
}=0,n_{\downarrow }=1)$. We then introduce the constriction which prevents
double occupancy in the quantum dot by means of Lagrange multiplier $\lambda
.$ The resulting model can be solved within the mean-field approach. Two
constants $b$ and $\lambda $ can be calculated from the following
self-consistent equations
\begin{equation}
-i\int \frac{dE}{2\pi }TrG_{0}^{<}(E)+b^{2}=1,
\end{equation}%
\begin{equation}
\lambda b^{2}=i\int \frac{dE}{2\pi }\sum_{\alpha }Tr\{G_{0}^{r}(E)\Sigma
_{\alpha }^{<}(E)+G_{0}^{<}(E)\Sigma _{\alpha }^{a}(E)\}.
\end{equation}%
Having constants $b$, $\lambda $ and using Eqs.(7-13), we can obtain the
full Green's function and finally give the charge and spin current in the
strong coupling limit. Fig.5 shows the Josephson current versus the two
superconducting phase difference $\Phi $ of two superconductors for the
various polarized laser frequencies $\omega .$ It is found that as the
frequency $\omega $ increases, the system will go from a $\pi $-state to a $%
0 $-state. This means that the strong Coulomb interaction will not affect
the dynamic $0-\pi $ transition.

In summary, we have presented a charge and spin pumping theory of quantum
dot coupled to two superconducting leads. The $\sigma _{-}$ polarized laser
field serving as the pumping forces can induce the dynamic $0-\pi $
transition and spin current. In contrast to the Josephson supercurrent, the
pumped spin current is an even function of the phase difference between two
superconductors. The strong Coulomb interaction in the quantum dot does not
change the dynamic $0-\pi $ transition behavior. Since the system under
investigation is within the reach of the present nano-technology, we hope
that the present theory can stimulate the further study of cavity
superconducting electronics and spintronics.

\begin{acknowledgments}
This work was supported by the National Science Foundation of China under
Grant No. 90303011, 10474034 and 60390070. B. Wang was also supported by
National Basic Research Program of China through Grant No. 2004CB619305 and
NCET-04-0462.
\end{acknowledgments}

Figure Captions

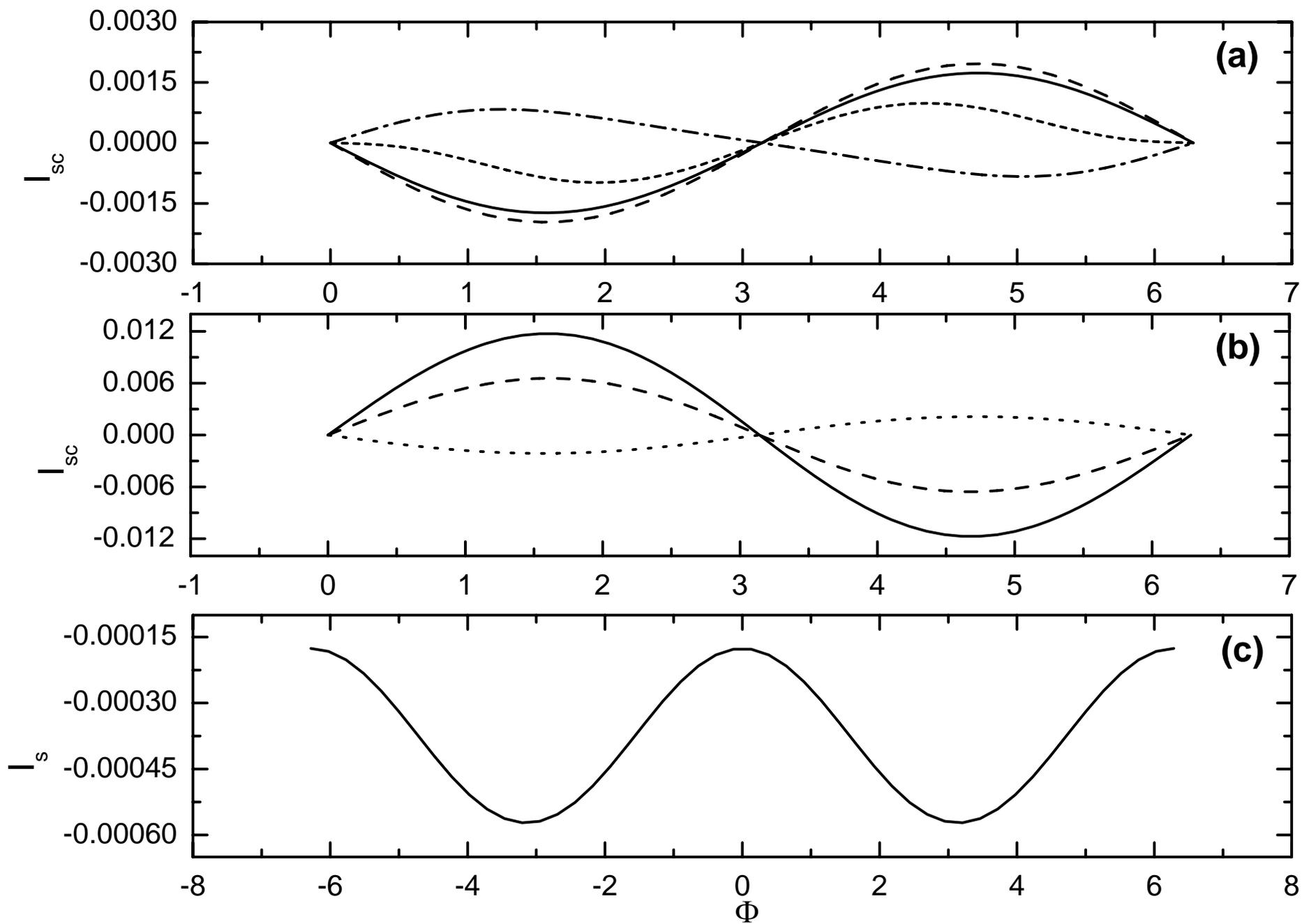
Fig.1. (a) Josephson current-phase relation for various laser frequencies $%
\omega .$ $\omega =0,$ $0.5,$ $1.5,$ $2.5$ correspond to solid, dash,
dash-dot, and short-dash lines, $\epsilon _{\uparrow }=\epsilon _{\downarrow
}=0,$ $r=0.6,$ $\Gamma _{L}=\Gamma _{R}=0.1.$ (b) Josephson current-phase
relation for various laser strength $r.$ $r=0.1,$ $0.3,$ $0.5$ correspond to
solid, dash, dot lines, $\epsilon _{\uparrow }=-0.2,$ $\epsilon _{\downarrow
}=-0.4,$ $\omega =0.3,$ $\Gamma _{L}=\Gamma _{R}=0.1.$ (c) The pumped spin
current versus the phase difference $\Phi $ between two superconctors. We
have set $\omega =2.5$ and other parameters are the same as those in
Fig.1(a). \ Note that all parameters are in unit of superconducting energy
gap $\Delta $.

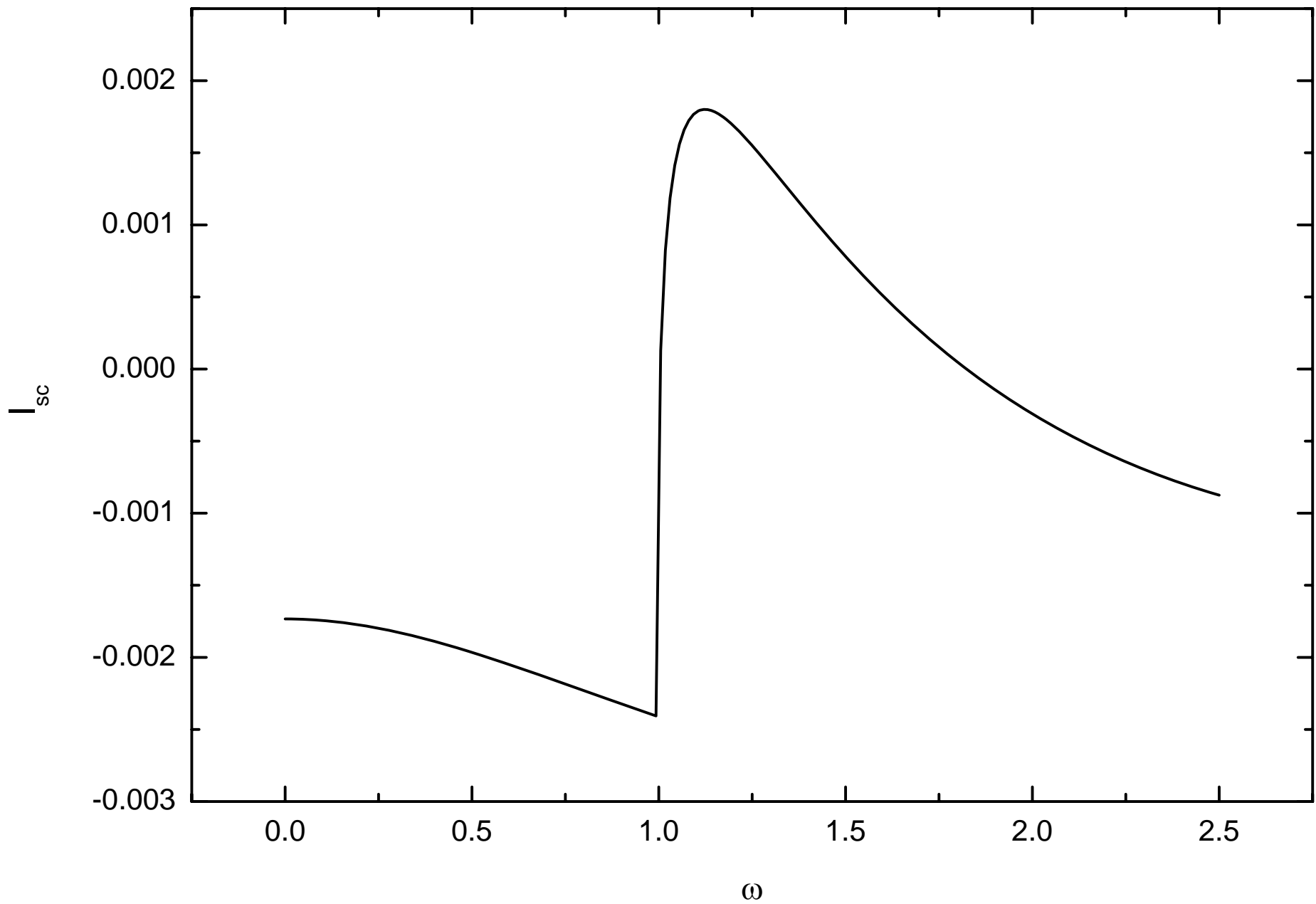
Fig.2. Josephson current $I_{sc}$ versus frequency $\omega $. Here $\Phi
=\pi /2$ and other parameters are the same as in Fig.1(a).

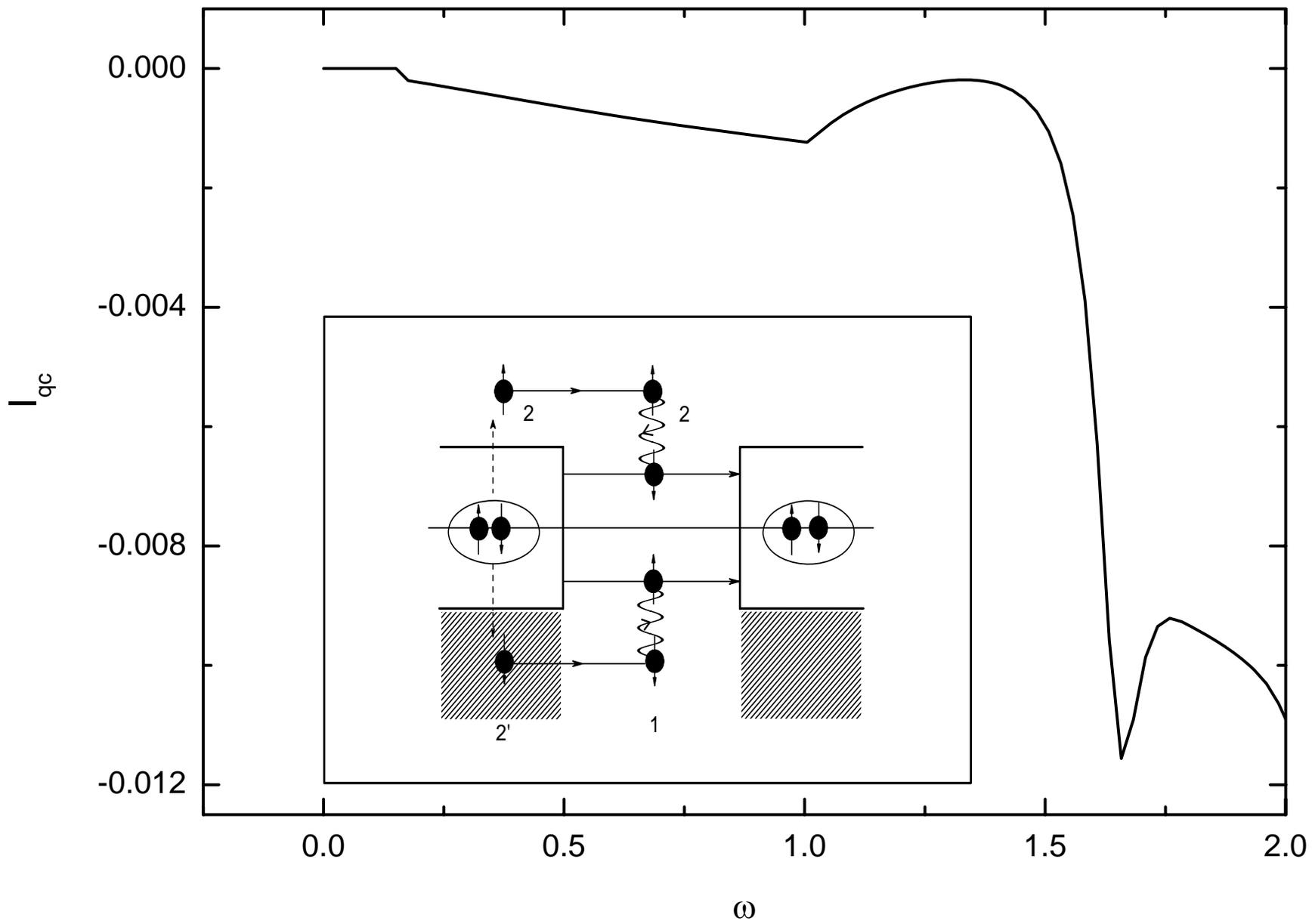
Fig.3. Pumped quasi-particle charge current $I_{qc}$ as a function of $%
\omega $. Here $\epsilon _{\uparrow }=0.9,$ $\epsilon _{\downarrow }=0.5,$ $%
r=0.6$ and other parameters are the same as in Fig.2(a). Inserted:
co-tunneling process for the pumped quasi-particle charge current.

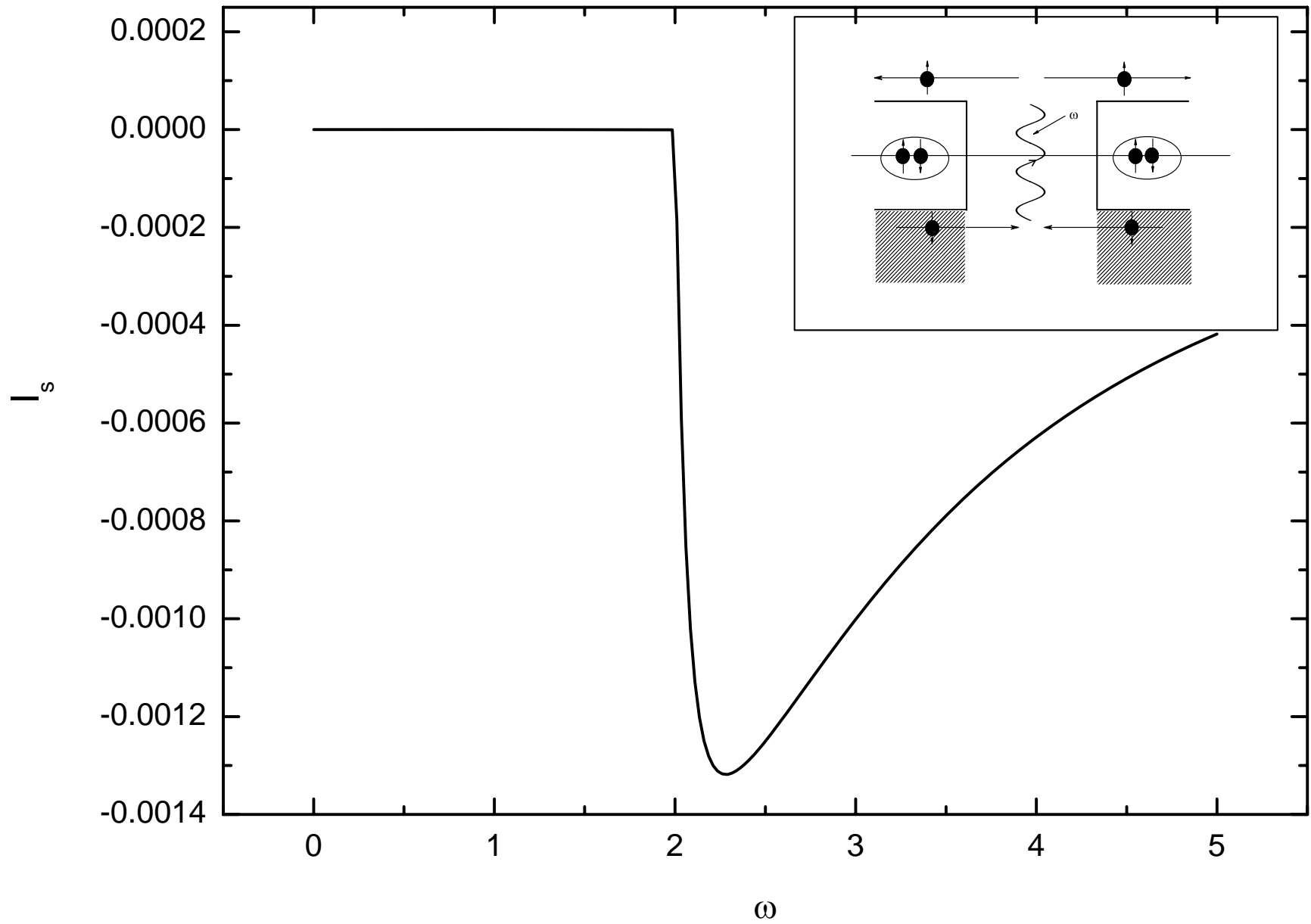
Fig.4. Pumped quasi-particle spin current $I_{s}$ as a function of $\omega $%
. Parameters are the same as in Fig.3. Inserted: spin-flip process for the
pumped spin current.

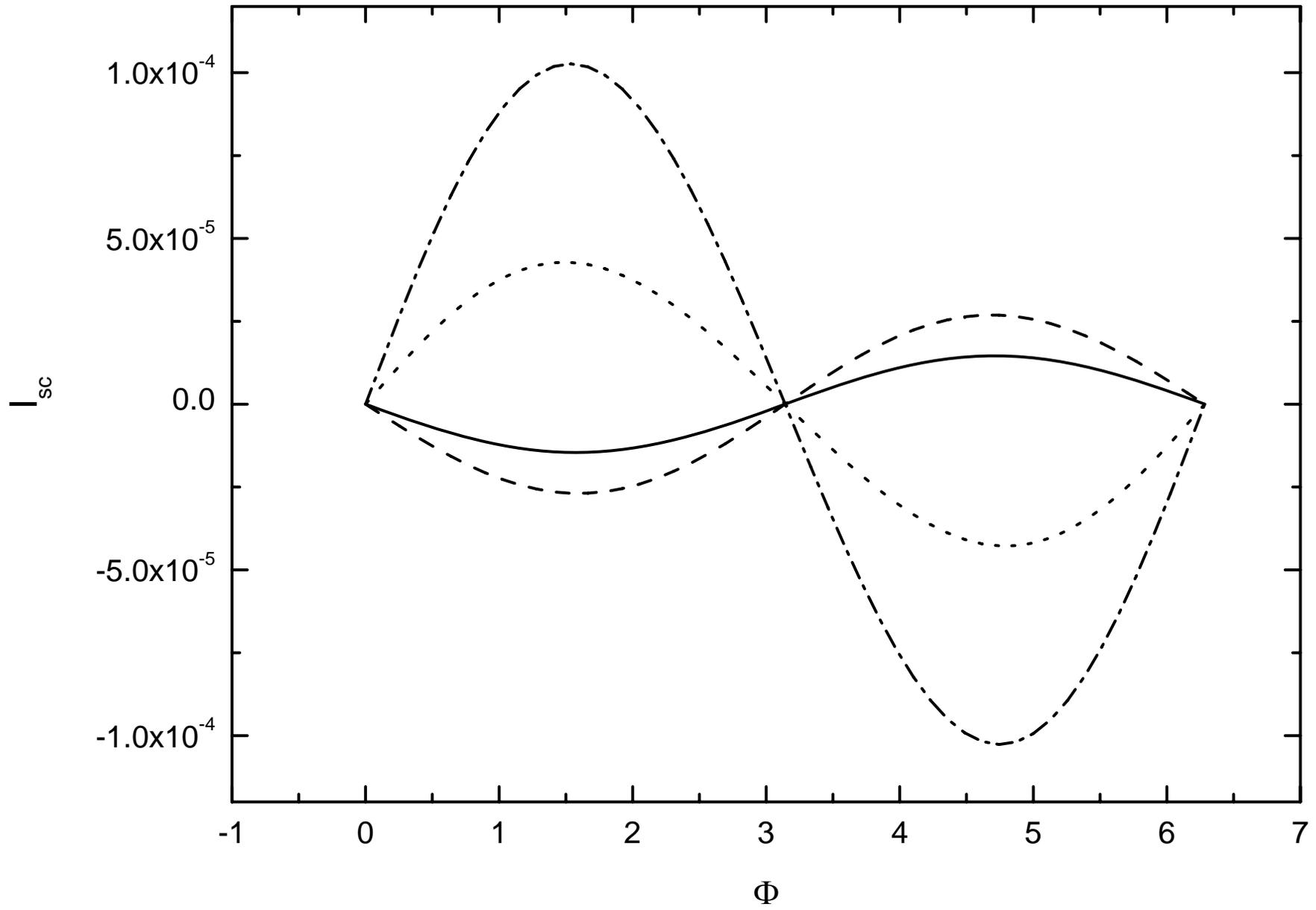
Fig.5. Josephson current-phase relation for various laser frequencies $%
\omega $ with a strong Coulomb interaction in the quantum dot. $\omega =0.6,$
$1.2,$ $2.2,$ $2.8$ correspond to solid, dash, dot, dash-dot line. Other
parameters are $\epsilon =-2,$ $\Gamma _{L}=\Gamma _{R}=0.5,$ $r=2,$ $\Phi
\equiv \pi /2,$ $\Delta =0.002,$ $W=100,$ $\beta =500.$


\begin{thebibliography}{99}
\bibitem{footnote1} The integral of the variable $X$ can be performed easily
since there exists $\delta $ function in the retarded Green's function.
\end{thebibliography}
\end{document}